\documentclass[doublecol]{epl2} 

 \renewcommand{\Im}{\mathop{\mathrm{Im}}\nolimits}  

\title{Macroscopic quantum dynamics of toroidal moment \\ in Ising-type rare-earth clusters}
\shorttitle{Macroscopic quantum dynamics of toroidal moment in Ising-type rare-earth clusters} 

\author{D. I. Plokhov\inst{1}\thanks{E-mail: \email{dmitry.plokhov@gmail.com}} \and A. I. Popov\inst{2} \and A. K. Zvezdin\inst{1}\thanks{E-mail: \email{zvezdin@gmail.com}}}
\shortauthor{D. I. Plokhov, A. I. Popov, A. K. Zvezdin}

\institute{ \inst{1} A. M. Prokhorov General Physics Institute of Russian Academy of Sciences, 38 Vavilov Str. 119991 Moscow, Russia \\
            \inst{2} Moscow Institute of Electronic Technology, 5 Pas. 4806, 124498 Zelenograd, Moscow, Russia }


\abstract{We study the quantum dynamics of polygonal rare-earth molecular clusters with Ising-type ion magnetization. It is shown that the ground state of such systems is a non-magnetic quasi-doublet of states with oppositely twisted ion spins. The states differ in sign of toroidal moment, which is a natural physical quantity to characterize the spin chirality of the clusters. The possibility of macroscopic quantum tunneling of toroidal moment between the states is predicted. The effects of an external current is considered, both in equilibrium and in the frames of the Landau-Zener-St\"uckelberg tunneling model. The special treatment is given for the most important case of triangular rare-earth clusters.}

\begin{document}

\maketitle \frenchspacing

\section{Introduction}

Magnetic nanoclusters are in the focus of intense research because they exhibit remarkable quantum properties such as macroscopic quantum tunneling (MQT) of magnetization, molecular bistability, quantum interference, Berry phase effects, etc. Besides the obvious fundamental significance, their investigation is of great importance for the fronts of nanoelectronics, namely molecular spintronics \cite{Boga,Sanv,Meie,Sonc}, as well as for full-scale quantum computer development in the aspect of overcoming of non-scalability and difficulties in qubit state control.

In this respect, magnetic molecular nanoclusters with spin chirality (Cu$_3$, V$_{15}$, Dy$_3$, etc.) are of special interest. The extra degree of freedom can be used for qubit coding. It was shown by the example of Cu$_3$ magnetic cluster \cite{Loss} that the chirality is the source of a spin-electric effect whereby the spin states of the cluster is driven by an electric field, which is better localized than magnetic one \cite{Nowa}. 

The alternative possibility is the clusters with the non-magnetic ground state. It is pointed \cite{Mila} that getting rid of the spin makes the qubit very little sensitive to magnetic noise, but creates at the same time a potential problem for measurement since there is no Zeeman coupling of the qubit to an external magnetic field. One of possible solutions to this problem relies on the presence of an orbital moment associated to the chirality \cite{Mila}.

Recent experimental and theoretical studies \cite{Luzo,Chib,EPLT} of a dysprosium based triangular cluster demonstrate the possibility of chirality control with an electric current (or just with crossed electric and magnetic fields) via interaction with toroidal moment, which is a natural characteristic of chirality \cite{EPLT}. It is remarkable that the cluster is of zero magnetic moment in the ground state in spite of the fact that there is a certain order in spin arrangement in it. The molecules with toroidal moment were first implied in \cite{Ceul}. Recently, such a molecule (namely Dy$_3$) was synthesized \cite{Tang} and experimentally investigated \cite{Luzo, Chib}. Other examples of molecules with toroidal moment are nanocluster V$_{15}$ \cite{V-15} and probably nanocluster Cu$_3$ \cite{Loss}. Toroidal ordering in crystals is reviewed in \cite{Kopa}.

Today, the quantum properties of chiral molecular clusters with toroidal moment is lacking of systematic research. In the present paper, the intriguing question of macroscopic quantum tunneling of toroidal moment (spin chirality) is considered for the first time. We predict the existence of Rabi oscillations in the clusters, which provides the good evidence for the MQT. The current driven dynamics is also considered, both in equilibrium and in the frames of the Landau-Zener-St\"uckelberg tunneling model. Due to the coupling between the toroidal moment and the current, it is possible to distinguish between states with opposite chirality, which is important for the purpose of qubit coding.

\section{The model}

We consider a spin ring, a system of $N$ non-Kramers rare-earth ions located in the apices of a regular polygon. The Hamiltonian of the rare-earth polygonal cluster reads as follows
	\begin{equation} \label{hamt}
		{\cal H} = \sum_{i=1}^{N} {\cal H}_{CF}^{(i)} + {\cal H}_{INT} + g_J \mu_B \sum_{i=1}^{N} {\bf H} {\bf J}_i, \\
	\end{equation}
where ${\cal H}_{CF}^{(i)}$ is the operator of the crystal field at the $i$-th ion location, ${\bf J}_i$ is the total angular momentum of the $i$-th ion, ${\bf H}$ is the external magnetic field strength, and ${\cal H}_{INT}$ is the Hamiltonian of the dipole and exchange interactions of the rare-earth ions in the cluster.

\begin{figure} \centering \includegraphics[scale=0.700]{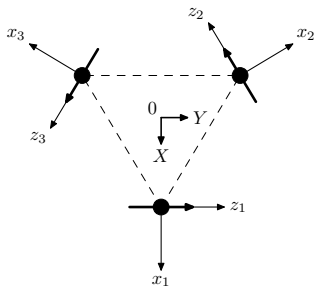} \caption{\label{spin} The spin structure of a triangular rare-earth molecular cluster and the local easy axes orientation in respect of the laboratory $XYZ$ reference frame. The thicker arrows represent the spins of the rare-earth ions in the molecule in the state with toroidal moment $T_Z = + 3 T_0$.} \end{figure}

The ground state of rare-earth ions in compounds is formed mainly due to the influence of a crystal field and often has a strong anisotropy of the magnetic moment. For example, the ground state of Dy$^{3+}$ ions in Dy$_3$ cluster is very close Kramers doublet $| M_J = \pm 15/2 \rangle$ \cite{Luzo,Chib,EPLT} and responds only to the $z_i$ local component of an external magnetic field (see fig. 1). The first excited state is separated from the ground one by the energy of 200~cm$^{-1}$. The wave functions of the excited state are close to $| M_J = \pm 13/2 \rangle$. We can therefore conclude that the environment of rare-earth ions is almost axially symmetric. In the case of non-Kramers ions, small asymmetrical perturbations remove degeneration, thus making the ground state be quasi-doublet, i.e. the two close singlets with splitting small compared with distance from the higher levels. According to the Griffiths theorem \cite{Grif}, the magnetic moments of the ions can be directed only along a specific axis, namely the local $z_i$-axes.
 
Let $| a_i \rangle$ and $| b_i \rangle$ be the eigenstates of the $i$-th (hereafter $i = 1, ..., N$) ion in the crystal field in respect of the local axes. The $z_i$-axes are perpendicular to the bisectors of the $N$-gon. The abscissa axes $x_i$ are then directed along the bisectors at the apices of the $N$-gon. Projection of the Hamiltonian in eq.~(\ref{hamt}) on the Hilbert space with functions $| \chi^{(i)}_{\pm} \rangle = (| a_i \rangle \pm i | b_i \rangle) / \sqrt{2}$ constituting the basic set yields the effective Hamiltonian
	\begin{equation} \label{heff}
		{\cal H}_{EFF} = - \frac{j}{2} \sum_{i \ne k}^{N} \sigma_{iz} \sigma_{kz} 
			- \frac{\Delta}{2} \sum_{i = 1}^{N} \sigma_{ix}
			- \sum_{i = 1}^{N} \tilde{\mu_i} H_{z_i} \sigma_{iz},
	\end{equation}
where $\sigma_x$, $\sigma_y$, and $\sigma_z$ stand for the Pauli matrices, $j$ is the exchange interaction constant, $\Delta$ is the energy gap between the singlet levels $| a_i \rangle$ and $| b_i \rangle$ (as a rule $\Delta << j$), and $\tilde{\mu_i} = g_J \mu_B \Im \langle a_i | J_{z_i} | b_i \rangle$. Usually, $| \chi_{\pm}^{(i)} \rangle \approx | M_J = \pm J \rangle \equiv | \pm \rangle$. It will be shown below that eq.~(\ref{heff}) is valid not only for non-Kramers ions, but for Kramers ions as well. In the latter case, the splitting $\Delta$ can be produced by an external magnetic field $H$, and $\Delta \rightarrow 0$ at $H \rightarrow 0$.

There are $2^N$ possible orderings of the ion spins in the rare-earth polygonal cluster. The wave functions of the cluster can be written as 
	\begin{equation} \label{ords}
		\chi_n = \prod_{i=1}^{N} | \sigma_{n_i} \rangle,
	\end{equation}
where $\sigma_{n_i}$ stands for the sign ("$+$" or "$-$") of the $i$-th ion ($i = 1, ..., N$) spin projection onto the local $z_i$ axis in the $n$-th state ($n = 1, ..., 2^N$).

The spin orderings can be characterized in terms of spin chirality. It is clear that spins in the states $\chi_n$ ($n = 1, ..., 2^{N-1}$) and their conjugates $\chi_n$ ($n = 2^{N-1} + 1, ..., 2^N$) are inversely twisted, i.e. the states have opposite chirality. The natural physical quantity associated with spin chirality in this case is the $T$-odd polar vector of the anapole (toroidal) moment, which corresponds to the first term of the toroidal family in the multipole expansion of an arbitrary electric current distribution. We would remind that the toroidal moment operator in the case of localized magnetic ions can be defined as $${\bf T} = \frac{1}{2} g_J \mu_B \sum_{i=1}^{N} \left[ {\bf r}_i \times {\bf J}_i \right],$$ where ${\bf r}_i$ is the radius-vector connecting the center of the $N$-gon with the $i$-th apex (all $r_i = r_0$).

Hereafter we deal with the dimensionless values of the toroidal moment $\tau = T_Z / T_0$, where $T_0 = \frac{1}{2} g_J \mu_B r_0 J$. Obviously, the toroidal moment of the system in a $\chi_n$ state is $\tau = q_{+} - q_{-}$, where $q_{\pm}$ is the number of "$+$"s and "$-$"s in the $\chi_n$ state. The values of the toroidal moments for conjugate states are different in sign.

\section{Macroscopic quantum tunneling of the toroidal moment}

In the limiting case $\Delta \rightarrow 0$ (i.e. $\Delta << j$) the ground state is $| + ... + \rangle$ and $| - ... - \rangle$, which is doubly degenerated. Such a situation brings up the actual question on the macroscopic quantum coherence (MQC), i.e. the cluster ground state degeneration removal, and on the macroscopic quantum tunneling (MQT) of the toroidal moment, for instance, from the state $| T_Z = + T \rangle$ to the state $| T_Z = - T \rangle$, where $T = N T_0 + {\cal O}((\Delta/j)^2)$ is the value of the toroidal moment at the finite quantity $\Delta$. Both questions come to calculation of the imaginary-time ($\tau = it$, $\tau \in [ 0; \tau_0 ]$) transition amplitude between the states, which can be treated in terms of the path integral \cite{Feyn}
	$$\left\langle + T \left| \exp \left( \frac{{\cal H} \tau_0}{\hbar} \right) \right| - T \right\rangle
		= \int \exp \left( - \frac{S_E}{\hbar}  \right) D\theta D\varphi,$$
where $D\theta D\varphi$ is the integrating measure, $$S_E = - \int_{0}^{\tau_0} {\cal L} (-i\tau) d\tau$$ is the Euclidean action, and ${\cal L}$ is the Lagrangian of the system. The wave functions of the rare-earth polygonal cluster in the considered Hilbert space, which is the direct product of the state spaces of each of the $N$ cluster ions, can be presented as
	\begin{equation} \label{psif}
		| \psi \rangle = \prod_{i=1}^{N}
		\left( \alpha(\theta_i,\varphi_i) | \chi_{+}^{(i)} \rangle + \beta(\theta_i,\varphi_i) | \chi_{-}^{(i)} \rangle \right),
	\end{equation}
where $\alpha(\theta,\varphi) = \cos \frac{\theta}{2}$ and $\beta(\theta,\varphi) = \sin \frac{\theta}{2} \cdot e^{-i\varphi}$ are the Cayley-Klein parameters depending on polar $\theta_i$ and azimuthal $\varphi_i$ angles of the ${\bf \sigma}_i$ in the
$i$-th local coordinate system with polar axis directed along the $i$-th easy axis.

The quantum dynamical equation of motion for the toroidal moment reads
	\begin{equation} \label{dtdt}
		i \hbar \frac{d {\bf T}}{d t} = \left[ {\bf T}, {\cal H} \right],
	\end{equation}
where the Hamilton operator ${\cal H}$ is defined by eq.~(\ref{hamt}). We suppose that the dynamics of ${\bf T} (t)$ is mainly conditional on the spins of the cluster ions, i.e. ${\bf r}_i (t) = const$. If one averages out eq.~(\ref{dtdt}) with the wave function from eq.~(\ref{psif}), one obtains
	\begin{equation} \label{dndt}
		\frac{\hbar}{2} \frac{d {\bf n}_i}{d t} = \left[ {\bf n}_i \times \frac{\partial E}{\partial {\bf n}_i} \right],
	\end{equation}	
where ${\bf n}_i = \{ \sin\theta_i\cos\varphi_i; \sin\theta_i\sin\varphi_i; \cos\theta_i \}$ are the unit vectors and 
	\begin{equation}
		E = \langle {\cal H} \rangle = - \frac{\Delta}{2} \sum_{i=1}^{N} n_{x_i} - \frac{j}{2} \sum_{i \ne k}^{N} n_{z_i} n_{z_k}
	\end{equation} is the energy of the system at the zero temperature.

It is crucial for the MQT to take place that there are coherent macroscopic processes, which naturally means that $\theta_1 = ... = \theta_N \equiv \theta$ and $\varphi_1 = ... = \varphi_N \equiv \varphi$. The symmetry of the system is not affected unless the external magnetic field has a component directed in the plane of the rare-earth polygon. In terms of the $(\theta, \varphi)$ variable set eqs.~(\ref{dndt}) take the form
	\begin{eqnarray} \label{eksy}
			\hbar \dot{\varphi} = 4 j \cos \theta - \Delta \cot \theta \cos \varphi, \\ \nonumber
			\hbar \dot{\theta} = - \Delta \sin \varphi,
	\end{eqnarray}
which are the Lagrange-Euler equations of the Lagrangian
	\begin{equation} \label{lagr}
		{\cal L} = \frac{N \hbar}{2} (1 - \cos \theta) \dot{\varphi} + N j \cos^2\theta + \frac{N \Delta}{2} \sin\theta \cos\varphi.
	\end{equation}

The first term in eq.~(\ref{lagr}) is the Wess-Zumino one. Generally, the term is calculated making use of well-known formula for the Berry phase and interpreted as a contribution conditional on the gauge field of the Dirac monopole located at the point ${\bf n} = (0; 0; 0)$. It is interesting that the Isingness of the considered system do not have an impact on the form of the Wess-Zumino term, i.e. on the interaction between the molecule and the geometry field of the Dirac monopole, as opposed to its interaction with the external field.

The probability of the tunneling per time unit in the quasi-classical limit is determined \cite{Legg} by the expression	
	$$P = A \exp(-B),$$ 
where $B = 2 S_E / \hbar$ is the Gamov constant, and 
	$$S_E = \int \left( -i \frac{3\hbar}{2} (1 - \cos \theta) \frac{d\varphi}{d\tau} + E(\theta, \varphi) \right) d\tau$$ 
is the Euclidean action calculated on the instanton tunneling trajectory. The pre-exponential factor $A \sim \omega_0$ \cite{Chud}, where $\omega_0$ is the instanton frequency (see below).

In order to calculate the tunneling trajectory and the corresponding contribution to the action we switch analytically in eq.~(\ref{eksy}) and eq.~(\ref{lagr}) to the imaginary time $\tau = i t$, and note that eqs.~(\ref{eksy}) have the first integral
	\begin{equation} \label{imot}
		j \cos^2\theta + \frac{\Delta}{2} \sin\theta \cos\varphi = const.
	\end{equation}

Now we consider the stationary points and the transition solution ("bounce trajectory") of the eqs.~(\ref{eksy}). Points $\varphi = 0$ and $\cos\theta = \pm \sigma_0 = \pm \sqrt{1 - (\Delta / 4j)^2}$ are the stationary points of the system, which are actually the terminal points of the tunneling between states with toroidal moment $T_Z = \pm T = \pm N T_0 \cdot \sigma_0$. Point $\theta = \pi / 2$ and $\varphi = 0$ is the minimum point of the inverted energy.

Eliminating the $\varphi$-variable from the system of eqs.~(\ref{eksy}), we come to the instanton equation
	\begin{equation} \label{imag}
		\frac{d \cos \theta}{d \tau} = \frac{2 j}{\hbar} \left( \sigma_0^2 - \cos^2\theta \right).
	\end{equation}
Integrating the equation and taking into account the initial conditions $\cos \theta |_{\tau = \pm \infty} = \pm \sigma_0$, we get the instanton solution $$\cos \theta = \sigma_0 \cdot \tanh \left( \frac{2 j \sigma_0 \tau}{\hbar} \right),$$ whence it follows that the instanton frequency $\omega_0$ is $$\omega_0 = \frac{4 j \sigma_0}{\hbar}.$$ The time dependence of the $\varphi$-variable is then specified by the motion integral, see eq.~(\ref{imot}):
	$$\cos \varphi = \frac{2 - \sigma_0^2 - \cos^2 \theta}{2 \sqrt{1-\sigma_0^2} \sin\theta}.$$
It should be noted that Euclidean angle $\varphi$ takes on imaginary values. Substituting the obtained explicit dependencies $\theta = \theta(\tau)$ and $\varphi = \varphi(\tau)$ into eq.~(\ref{lagr}) for the system Lagrangian and calculating the action $S_E$ on the tunneling trajectory, we obtain for the Gamov constant
	\begin{equation} \nonumber
		B = \frac{2 S_E}{\hbar} = \frac{2 N j}{\hbar} \int_{-\infty}^{+\infty}
		\frac{\cos^2\theta \cdot (\sigma_0^2 - \cos^2\theta)}{1 - \cos^2\theta} d\tau,
	\end{equation}
which finally comes to
	\begin{equation} \label{gamc}
		B = N \left( \ln \frac{1+\sigma_0}{1-\sigma_0} - 2 \sigma_0 \right). 
	\end{equation}

It is seen that $B \cong 2 N \sigma_0^3 / 3$ if $\sigma_0 \rightarrow 0$, and also $B \rightarrow \infty$ if $\sigma_0 \rightarrow 1$. The tunneling frequency is $\nu \sim \frac{\omega_0}{\pi} e^{-B}$. It is determined by the $\sigma_0$, the value of the effective toroidal moment. If the exchange interaction between the cluster ions is strong enough ($j >> \Delta$), then $\sigma_0 \rightarrow 1$, and the tunneling probability vanishes.

The quantum mechanism of the tunneling dominates over thermal-activated processes of the spin reorientation at low enough temperatures. The temperature dependence of the speed of the latter is $\sim \exp(-U_0/kT)$, where $U_0$ is the height of the energy barrier. The crossover temperature $T^{\ast}$, at which both factors are of equal significance is defined by the condition $B = U_0 / kT$, therefore
	\begin{equation} \label{temc}
		T^{\ast} = \frac{2 N j}{k B(\sigma_0)} \cdot
		\left( 1 - \frac{\sigma_0^2}{2} - \sqrt{1 - \sigma_0^2} \right).
	\end{equation}
It follows from the eq.~(\ref{temc}) that $T^{\ast} \rightarrow 0$ if $\sigma_0 \rightarrow 0$.

\section{Rare-earth triangular clusters}
		
The archetype of the noncollinear Ising model is a recently synthesized molecular dysprosium triangle \cite{Luzo,Tang}. The standard, based on the path integral, quasiclassical technique to handle a problem of macroscopic quantum tunneling \cite{Feyn, Hemm, Chut, Menz, Legg} is only qualitatively suitable for rare-earth triangles due to low spin of the system. To get quantitative results we have to deal with the solutions of the Schroedinger equation in the relative Hilbert space.


There are eight possible orderings of the ion spins in the rare-earth triangular cluster, which are $\chi_1 = | + + + \rangle$, $\chi_2 = | - + \: + \rangle$, $\chi_3 = | + - \: + \rangle$, $\chi_4 = | + + \: - \rangle$, and $\chi_n = \chi_{n - 4} (+ \leftrightarrow -)$ for $n = 5, 6, 7, 8$, see eq.~(\ref{ords}).

In the basis of the eight vectors 
	$$ 
		\begin{array}{cc} 
			\psi_1 = \chi_1, & \psi_2 = (\chi_2 + \chi_3 + \chi_4) / \sqrt{3}, \\
			\psi_3 = (2\chi_2 - \chi_3 - \chi_4) / \sqrt{6}, & \psi_4 = (\chi_3 - \chi_4) / \sqrt{2}, 
		\end{array}
	$$
($\psi_{n+4}$ are obtained as $\psi_n$ by the replacement $\chi_n \rightarrow \chi_{n+4}$ for $n = 1, 2, 3, 4$), the matrix of the Hamiltonian in eq.~(\ref{heff}) in a zero magnetic field reads 
	\begin{equation} \label{8xx8}
		{\cal H} = \left( \begin{array}{cc}
		{\cal H}_0        & {\cal H}_{\Delta} \\
		{\cal H}_{\Delta} &        {\cal H}_0 
		\end{array} \right),
	\end{equation}
where 
	$${\cal H}_0 = \left( \begin{array}{cccc} 
	              -3j/4 & -\Delta\sqrt{3}/2 &        0 &        0 \\
	  -\Delta\sqrt{3}/2 &               j/4 &        0 &        0 \\
                    0 &                 0 &      j/4 &        0 \\
                    0 &                 0 &        0 &      j/4 
    \end{array} \right),$$
and  
	$${\cal H}_{\Delta} = \left( \begin{array}{cccc} 
		0 &       0 &        0 &        0 \\
	  0 & -\Delta &        0 &        0 \\
    0 &       0 & \Delta/2 &        0 \\
    0 &       0 &        0 & \Delta/2 
    \end{array} \right).$$

It is possible to obtain the eigenvalues and eigenstates of the Hamiltonian, ${\cal H} \Phi_n = E_n \Phi_n$ ($n = 1, ..., 8$). The relevant explicit expressions are given in table~\ref{eigs}. Clearly, the eigenvectors $\Phi_n$ ($n = 1,
..., 8$) are an orthonormal set.

\begin{largetable} \caption{\label{eigs} The eigenenergies $\varepsilon_n = E_n / j$ and the eigenstates $\Phi_n$ ($n = 1, ..., 8$) of the Hamiltonian in eq.~(\ref{8xx8}), the following designations are used: $\varepsilon_{\pm} (x) = - \frac{1-2x}{4} \pm
\frac{d(x)}{2}$ and $f_{\pm}(x) = \frac{1}{2} \sqrt{1 \pm \frac{1+x}{d(x)}}$, where $d(x) = \sqrt{1+2x+4x^2}$ and $x = \Delta / j$.}
	\begin{center} \begin{tabular}{|c|c|c|c|c|c|c|c|}
  	\hline
  	1 & 2 & 3 & 4 & 5 & 6 & 7 & 8 \\
  	\hline
  	$\varepsilon_{+}(-x)$ & $\varepsilon_{-}(-x)$ & $\varepsilon_{+}(x)$ & $\varepsilon_{-}(x)$ & 
  	$\frac{1 + 2x}{4}$ & $\frac{1 + 2x}{4}$ & $\frac{1 - 2x}{4}$ & $\frac{1 - 2x}{4}$ \\
    \hline
    $ f_{-}(-x)$ & $f_{+}(-x)$ & $ f_{-}(x)$ & $ f_{+}(x)$ &                     0 &                     0 & 0 &                     0 \\
    $-f_{+}(-x)$ & $f_{-}(-x)$ & $-f_{+}(x)$ & $ f_{-}(x)$ &                     0 &                     0 & 0 &                     0 \\
              0 &           0 &           0 &           0 & 0 & $\frac{1}{\sqrt{2}}$ & 0 & $\frac{1}{\sqrt{2}}$ \\
              0 &           0 &           0 &           0 & $\frac{1}{\sqrt{2}}$ &                     0 & $\frac{1}{\sqrt{2}}$ & 0 \\
    $ f_{-}(-x)$ & $f_{+}(-x)$ & $-f_{-}(x)$ & $-f_{+}(x)$ & 0 & 0 & 0 & 0 \\
    $-f_{-}(-x)$ & $f_{-}(-x)$ & $ f_{+}(x)$ & $-f_{-}(x)$ & 0 & 0 & 0 & 0 \\
              0 &         0 &             0 &           0 & 0 & $\frac{1}{\sqrt{2}}$ & 0 & $-\frac{1}{\sqrt{2}}$ \\
              0 &         0 &             0 &           0 & $\frac{1}{\sqrt{2}}$ & 0 & $-\frac{1}{\sqrt{2}}$ & 0 \\
    \hline
    \end{tabular} \end{center}
\end{largetable}

The eigenstate with the lowest energy is $\Phi_2$, which is the superposition of the states close to the states with $| \tau = \pm 3 \rangle$, i.e. $\Phi_2 = \left( \psi_{+} + \psi_{-} \right) / \sqrt{2}$, where we make use of $\psi_{\pm} = \sqrt{2} \left( f_{+} (-x) \cdot \psi_{1,5} + f_{-} (-x) \cdot \psi_{2,4} \right)$. The expected values of the toroidal moment in these states are 
	\begin{equation} \label{taux}
		\langle \psi_{\pm} | \hat{\tau} | \psi_{\pm} \rangle = \pm \left( 2 + \frac{1-x}{\sqrt{1-2x+4x^2}} \right).
	\end{equation} 

Supposed that initially the system is in the $| \psi_{+} \rangle$-state, one can obtain the probability $P(t)$ to find the system in the $| \psi_{-} \rangle$-state by the time moment of $t$ 
	\begin{eqnarray} \label{rabi} 
		P(t) = \left| \langle \psi_{-} | \psi_{t} \rangle \right|^2 = \nonumber \\
		= \left| \sum_{n=1}^{8} \langle \Phi_n | \psi_{-} \rangle \langle \psi_{+} | \Phi_n \rangle 
		\exp \left( -\frac{i}{\hbar} E_n t \right) \right|^2, 
	\end{eqnarray}
where $| \psi_t \rangle = \exp \left( - \frac{i}{\hbar} {\cal H} t \right) | \psi_{+} \rangle $.

\begin{figure} \centering \includegraphics[scale=0.301]{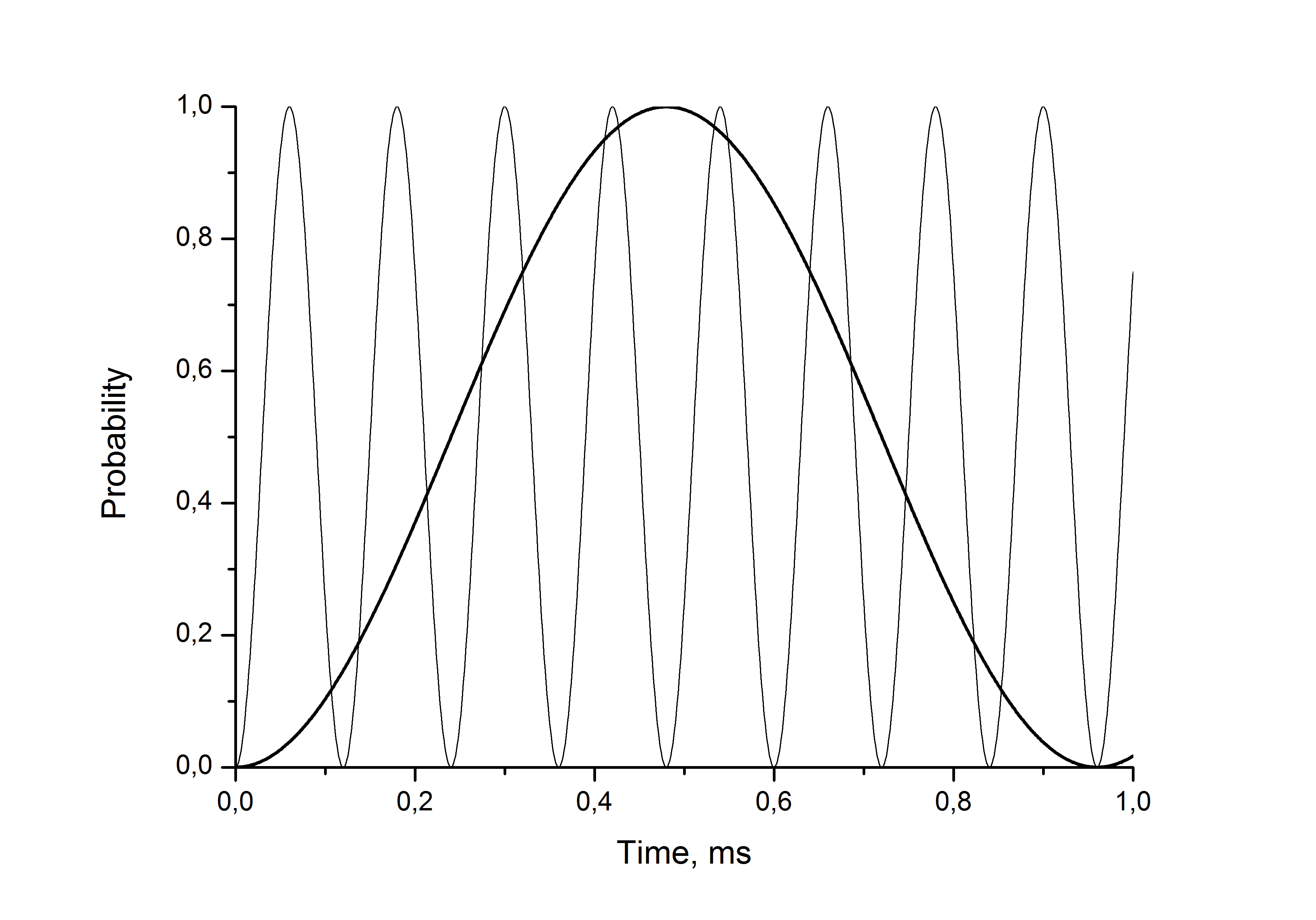} \caption{\label{oscs} The Rabi-type oscillations between states $| \tau \approx \pm 3 \rangle $, the longer period corresponding to $\Delta = 0.003 j$ spans over eight shorter periods corresponding to the doubled splitting $\Delta = 0.006 j$, thus showing the cubical oscillation period dependence on the splitting.} \end{figure}

The probability function $P(t)$ oscillates (see fig.~\ref{oscs}), thus giving the clear evidence of the macroscopic quantum tunneling of the toroidal moment in the system. If the level splitting $\Delta$ is small compared to the exchange constant $j$, the oscillation frequency depends on parameter $x = \Delta / j$ as $\nu = \alpha x^3$ with $\alpha = 48$ GHz, so for the typical value $\tau = \nu^{-1} \sim 1$ ms \cite{Luzo} we have $x \sim 0.003$, i.e. $\Delta \sim 0.03$ cm$^{-1}$ if $j = 10$ cm$^{-1}$. The expected value of the toroidal moment given by eq. (\ref{taux}) is very close to $\pm 3$, namely $\langle \hat{\tau} \rangle \approx \pm 3 (1 - x^2 / 2) \sim \pm (3 - 1 \cdot 10^{-5})$.

Let us now consider the influence of magnetic and electric fields on the MQT of the toroidal moment. The states of a rare-earth ion of the cluster in the crystal field are $| 1 \rangle \equiv | a \rangle$, $| 2 \rangle \equiv | b \rangle$, and $| k \rangle$, the energy levels are $-\Delta / 2$, $\Delta / 2$, and $W_k \sim 100$ cm$^{-1}$ respectively. In the general case, the projection of the Hamiltonian ${\cal V} = {\cal H}_{CF} + {\cal H}_Z$ (with the Zeeman part ${\cal H}_Z = g \mu_B {\bf H J}$) onto the quasidoublet states gives $\tilde{V}_{ij} = V_{ij} - \sum_k V_{ik} V_{kj} / W_k$, $i, j = 1, 2$.

In the case of the magnetic field applied perpendicular to the plane of the ion triangle 
	\begin{equation} \label{tvij}
		 \tilde{V}_{11} = -\frac{\Delta}{2} - q_1 H^2, \ \tilde{V}_{22} = \frac{\Delta}{2} - q_2 H^2, \ \tilde{V}_{12} = \tilde{V}_{21} = 0,
	\end{equation} 
where $q_{1,2} = g_J^2 \mu_B^2 \sum_k \frac{1}{W_k} \langle 1,2 | J_y | k \rangle \langle k | J_y | 1,2 \rangle$.

The symmetry of the system is not broken in the perpendicular field, which, as it follows from eq.~(\ref{tvij}), results in the renormalization of the splitting $\Delta$ only $$\Delta \rightarrow \tilde{\Delta} = \Delta + (q_1 - q_2) H^2.$$ 

Moreover, in the case of Kramers (dysprosium) ions, such a field produces a small splitting $\Delta \sim g_y H$ of the ground state due to the smallness of $g_y$, thus removing the degeneration. Since the typical value of relaxation time is 1 ms \cite{Luzo}, the splitting $\Delta$ is estimated at 0.03 cm$^{-1}$. The factor $g_y$ then equals to 0.06 in a 1 T magnetic field. This way, the theory developed is valid not only for non-Kramers but also for Kramers Ising-type ions as well. 

An external static electric field perpendicular to the plane of the ion triangle has no influence on the considered system, because the MEE properties of the rare-earth triangular clusters are caused by the field, directed in the plane of the ion triangle
\cite{EPLT}.

\section{The current driven dynamics}

Let us consider now the interaction between the toroidal moment and an external electric current. To describe the interaction we put the term $\hat{V} = \frac{4\pi}{c} {\bf j} \hat{{\bf T}}$ into the Hamiltonian in eq.~(\ref{8xx8}). The electric current ${\bf j}$ could be the displacement current ${\bf j} = \frac{1}{4\pi} \frac{\partial {\bf E}}{\partial t}$. If the electric field strength ${\bf E}$ linearly depends on the time and is directed along the $Z$-axis, then $\hat{V} = \frac{v}{c} \cdot T_0 \cdot \hat{\tau}$, where $v = \partial E_Z / \partial t$. The $\hat{\tau}$-operator matrix has the diagonal form $\tau = (+3,+1,+1,+1,-3,-1,-1,-1)$ in the basis of $\psi_n$ ($n = 1, ..., 8$). The Hamiltonian ${\cal H} + \hat{V}$ of the system can also be diagonalized,
the eigenvalues $E^{(j)}_n$ and eigenvectors $\Phi^{(j)}_n$ ($n = 1, ..., 8$) are not given here due to their unhandiness. The energies of the two low-lying levels could be approximated (for $x << 1$) in the vicinity of the avoided level crossing at $j_z = 0$ as
	\begin{equation} \label{lowe}
		E_{\pm} (x,j_z) = E_0 \mp \sqrt{\left( \frac{4\pi}{c} \cdot 3T_0 \cdot j_z \right)^2 + \left( \frac{\delta}{2} \right)^2},
	\end{equation}
where the crossing energy is $E_0 = - \frac{3}{4} j$ and the avoided level splitting is $\delta = \frac{3}{2} j x^3$. 

\begin{figure} \centering \includegraphics[scale=0.301]{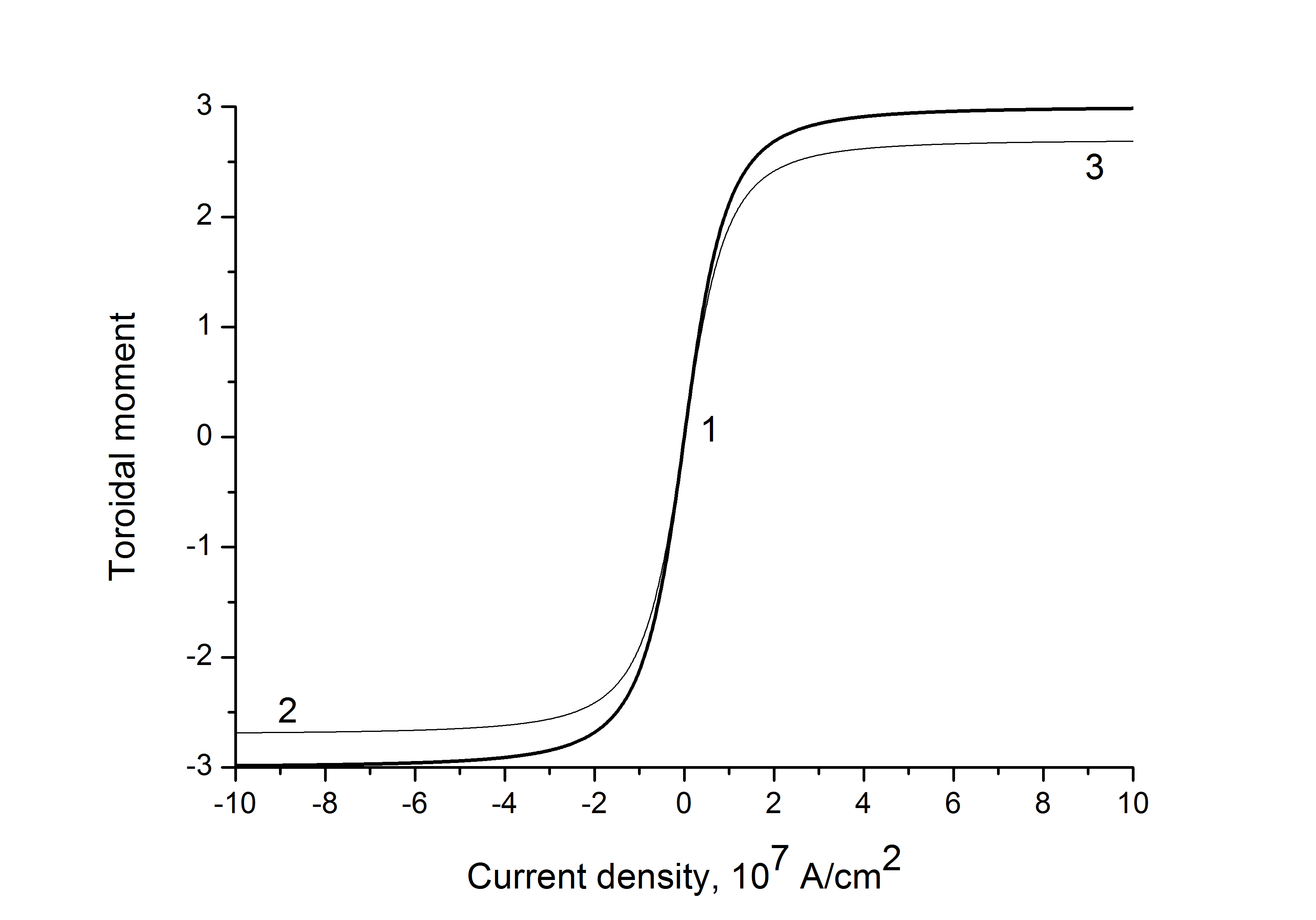} \caption{\label{tmeqlz} The zero temperature plot of toroidal (anapole) moment projection onto the laboratory $Z$-axis vs the external current density for $\Delta = 0.003 j$: thicker curve (1) in equilibrium, the other curves in the case of the finite value of current sweeping rate $10^{13}$ A/(cm$^2\cdot$s), (2) from $\tau = + 3$ to $\tau = - 2.7$ and (3) from $\tau = - 3$ to $\tau = + 2.7$.} \end{figure}

The toroidal moment is a conjugate variable against the electric current density and can be found at zero temperature as 
	\begin{equation}
		\tau (x) = \frac{c}{4 \pi T_0} \frac{\partial E_{-} (x,j_z)}{\partial j_z}.
	\end{equation}
The plot of the relative (equilibrium) dependence is given in fig.~\ref{tmeqlz} (curve 1). It is seen, that the current changes the direction of the spin twisting to the opposite, thus reversing the toroidal moment of the system in a way of the relatively sharp jump of $\Delta \tau = 6$. The anapole moment reversal, which could be called a reanapolization, requires the currents of $10^7$ A/cm$^2$, which seems to be quite achievable in experiment.

The tunneling processes when sweeping the current $j_z$ at a constant rate over an avoided energy level crossing can be treated in the frames of the Landau-Zener tunneling model. The probability $P$ to change the state characterized by the quantum number of the toroidal moment at the avoided level crossing is given by the expression
	\begin{equation} \label{laze}
		P = 1 - \exp \left( - \frac{\pi \delta^2}{(4\pi/c) \hbar \cdot T_0 \cdot |\tau_1 - \tau_2| \cdot |d j_z/d t|} \right),
	\end{equation}
where $\tau_1 = -3$ and $\tau_2 = +3$ are the toroidal moment quantum numbers of the avoided level crossing with the splitting $\delta = \frac{3}{2} j x^3$, and $d j_z/d t$ is the constant current sweeping rate. For the rate of $d j_z / d t \sim 10^{13}$ A/(cm$^2\cdot$s) and $x = 0.003$ we have $P \approx 0.95$.

With the Landau-Zener-St\"uckelberg model in mind, we can now start to understand qualitatively the hysteresis in the system considered (see fig.~\ref{tmeqlz}). Let us start at a sufficiently large negative current $j_z$. At very low temperature, all molecules are in the $| \tau = -3 \rangle$ ground state. When the current is come down to zero, all molecules will stay in the $| \tau = -3 \rangle$ ground state. 

When passing the current over the avoided level crossing region at $j_z \approx 0$, there is Landau-Zener probability $P$ ($0 < P < 1$) to tunnel from the $| \tau = -3 \rangle$ to the $| \tau = +3 \rangle$ state. The dependence of the toroidal moment on the current then reads as follows 
	\begin{equation}
		\tau (x,j_z) = P \cdot \frac{\partial E_{+}(x,j_z)}{\partial j_z} + (1 - P) \cdot \frac{\partial E_{-}(x,j_z)}{\partial j_z}
	\end{equation}
shown in fig.~\ref{tmeqlz} (curve 3). The toroidal moment undergoes the jump of $\Delta \tau = 6 P$. So the average value of toroidal moment $\tau$ becomes less than 3, namely $\tau = 2.7$ for $P = 0.95$. The relevant mixed quantum state relaxes to the equilibrium $| \tau = +3 \rangle$ state due to interaction with environment (the $3 \rightarrow 1$ process in fig.~\ref{tmeqlz}). 

Sweeping the current in the opposite direction from the $| \tau = +3 \rangle$ state, we arrive to the mixed state with toroidal moment average value of $-2.7$, see curve 2 in fig.~\ref{tmeqlz}. The state relaxes as above to the equilibrium $| \tau = -3 \rangle$ state (the $2 \rightarrow 1$ process in fig.~\ref{tmeqlz}). This way, we will come to the closed hysteresis loop.

\section{Conclusion} In the present work, it is shown for the first time that there exists a possibility of the macroscopic quantum tunneling of toroidal moment (spin chirality) in the molecular system based on Ising-type rare-earth ions. It is highly important that the ground state of such magnetically ordered systems has zero magnetic moment and should be characterized by toroidal moment. This is the reason why the slow relaxation observed experimentally in such systems cannot be caused by the MQT of the magnetic moment, but can possibly be accounted for the MQT of the toroidal moment instead.

\acknowledgments

We wish to acknowledge the financial support of the Russian Foundation for Basic Research (projects 08-02-01068, 09-02-11309, and 10-02-90475). One of the authors (A.K.Z.) thanks B. Barbara, A. Ceulemans, L.F. Chibotaru, and A. Soncini for discussions.

\end{document}